\documentclass[a4paper,11pt]{article}
\usepackage[T1]{fontenc} 

\makeatletter
\gdef\@fpheader{ }
\gdef\@journal{ }
\makeatother
\RequirePackage{amsmath}
\RequirePackage{amssymb}
\RequirePackage{epsfig}
\RequirePackage{graphicx}
\RequirePackage[numbers,sort&compress]{natbib}
\RequirePackage{color}
\RequirePackage[colorlinks=true
,urlcolor=blue
,anchorcolor=blue
,citecolor=blue
,filecolor=blue
,linkcolor=blue
,menucolor=blue
,pagecolor=blue
,linktocpage=true
,pdfproducer=medialab
,pdfa=true
]{hyperref}

\newif\ifnotoc\notocfalse
\newif\ifemailadd\emailaddfalse
\newif\iftoccontinuous\toccontinuousfalse
\makeatletter
\def\@subheader{\@empty}
\def\@keywords{\@empty}
\def\@abstract{\@empty}
\def\@xtum{\@empty}
\def\@dedicated{\@empty}
\def\@arxivnumber{\@empty}
\def\@collaboration{\@empty}
\def\@collaborationImg{\@empty}
\def\@proceeding{\@empty}
\def\@preprint{\@empty}

\newcommand{\subheader}[1]{\gdef\@subheader{#1}}
\newcommand{\keywords}[1]{\if!\@keywords!\gdef\@keywords{#1}\else%
\PackageWarningNoLine{\jname}{Keywords already defined.\MessageBreak Ignoring last definition.}\fi}
\renewcommand{\abstract}[1]{\gdef\@abstract{#1}}
\newcommand{\dedicated}[1]{\gdef\@dedicated{#1}}
\newcommand{\arxivnumber}[1]{\gdef\@arxivnumber{#1}}
\newcommand{\proceeding}[1]{\gdef\@proceeding{#1}}
\newcommand{\xtumfont}[1]{\textsc{#1}}
\newcommand{\correctionref}[3]{\gdef\@xtum{\xtumfont{#1} \href{#2}{#3}}}
\newcommand\jname{JHEP}
\newcommand\acknowledgments{\section*{Acknowledgments}}

\newcommand\preprint[1]{\gdef\@preprint{\hfill #1}}

\makeatother

\newcommand\note[2][]{%
\if!#1!%
\stepcounter{footnote}\footnotetext{#2}%
\else%
{\renewcommand\thefootnote{#1}%
\footnotetext{#2}}%
\fi}

\makeatletter
\newtoks\auth@toks
\renewcommand{\author}[2][]{%
  \if!#1!%
    \auth@toks=\expandafter{\the\auth@toks#2\ }%
  \else
    \auth@toks=\expandafter{\the\auth@toks#2$^{#1}$\ }%
  \fi
}
\makeatother
\makeatletter
\newtoks\affil@toks\newif\ifaffil\affilfalse
\newcommand{\affiliation}[2][]{%
\affiltrue
  \if!#1!%
    \affil@toks=\expandafter{\the\affil@toks{\item[]#2}}%
  \else
    \affil@toks=\expandafter{\the\affil@toks{\item[$^{#1}$]#2}}%
  \fi
}
\makeatother
\makeatletter
\newtoks\email@toks\newcounter{email@counter}%
\newcommand{\emailAdd}[1]{%
\emailaddtrue%
\ifnum\theemail@counter>0\email@toks=\expandafter{\the\email@toks, \@email{#1}}%
\else\email@toks=\expandafter{\the\email@toks\@email{#1}}%
\fi\stepcounter{email@counter}}
\newcommand{\@email}[1]{\href{mailto:#1}{\tt #1}}
\makeatother

\makeatletter
\newcommand*\collaboration[1]{\gdef\@collaboration{#1}}
\newcommand*\collaborationImg[2][]{\gdef\@collaborationImg{#2}}
\makeatletter
\newcommand\afterLogoSpace{\smallskip}
\newcommand\afterSubheaderSpace{\vskip3pt plus 2pt minus 1pt}
\newcommand\afterProceedingsSpace{\vskip21pt plus0.4fil minus15pt}
\newcommand\afterTitleSpace{\vskip23pt plus0.06fil minus13pt}
\newcommand\afterRuleSpace{\vskip23pt plus0.06fil minus13pt}
\newcommand\afterCollaborationSpace{\vskip3pt plus 2pt minus 1pt}
\newcommand\afterCollaborationImgSpace{\vskip3pt plus 2pt minus 1pt}
\newcommand\afterAuthorSpace{\vskip5pt plus4pt minus4pt}
\newcommand\afterAffiliationSpace{\vskip3pt plus3pt}
\newcommand\afterEmailSpace{\vskip16pt plus9pt minus10pt\filbreak}
\newcommand\afterXtumSpace{\par\bigskip}
\newcommand\afterAbstractSpace{\vskip16pt plus9pt minus13pt}
\newcommand\afterKeywordsSpace{\vskip16pt plus9pt minus13pt}
\newcommand\afterArxivSpace{\vskip3pt plus0.01fil minus10pt}
\newcommand\afterDedicatedSpace{\vskip0pt plus0.01fil}
\newcommand\afterTocSpace{\bigskip\medskip}
\newcommand\afterTocRuleSpace{\bigskip\bigskip}
\newlength{\affiliationsSep}

\DeclareFixedFont\trfont{OT1}{phv}{b}{sc}{11}

\renewcommand\maketitle{
\pagestyle{empty}
\thispagestyle{titlepage}
\setcounter{page}{0}
\noindent{\small\scshape\@fpheader}\@preprint\par

\afterLogoSpace
\if!\@subheader!\else\noindent{\trfont{\@subheader}}\fi
\afterSubheaderSpace
\if!\@proceeding!\else\noindent{\sc\@proceeding}\fi
\afterProceedingsSpace
{\LARGE\flushleft\sffamily\bfseries\@title\par}
\afterTitleSpace
\afterRuleSpace
\if!\@collaboration!\else
{\Large\bfseries\sffamily\raggedright\@collaboration}\par
\afterCollaborationSpace
\fi
\if!\@collaborationImg!\else
{\normalsize\bfseries\sffamily\raggedright\@collaborationImg}\par
\afterCollaborationImgSpace
\fi
{\bfseries\raggedright\sffamily\the\auth@toks\par}
\afterAuthorSpace
\ifaffil\begin{list}{}{%
\setlength{\leftmargin}{0.28cm}%
\setlength{\labelsep}{0pt}%
\setlength{\itemsep}{\affiliationsSep}%
\setlength{\topsep}{-\parskip}}
\itshape\small%
\the\affil@toks
\end{list}\fi
\afterAffiliationSpace
\ifemailadd 
\noindent\hspace{0.28cm}\begin{minipage}[l]{.9\textwidth}
\begin{flushleft}
\textit{E-mail:} \the\email@toks
\end{flushleft}
\end{minipage}
\else 
\PackageWarningNoLine{\jname}{E-mails are missing.\MessageBreak Plese use \protect\emailAdd\space macro to provide e-mails.}
\fi
\afterEmailSpace
\if!\@xtum!\else\noindent{\@xtum}\afterXtumSpace\fi
\if!\@abstract!\else\noindent{\renewcommand\baselinestretch{.9}\textsc{Abstract:}}\ \@abstract\afterAbstractSpace\fi
\if!\@keywords!\else\noindent{\textsc{Keywords:}} \@keywords\afterKeywordsSpace\fi
\if!\@arxivnumber!\else\noindent{\textsc{ArXiv ePrint:}} \href{http://arxiv.org/abs/\@arxivnumber}{\@arxivnumber}\afterArxivSpace\fi
\if!\@dedicated!\else\vbox{\small\it\raggedleft\@dedicated}\afterDedicatedSpace\fi
\ifnotoc\else
\iftoccontinuous\else\newpage\fi
\afterTocSpace
\afterTocRuleSpace
\fi
\setcounter{footnote}{0}
\pagestyle{myplain}\pagenumbering{arabic}
} 

\renewcommand{\baselinestretch}{1.1}\normalsize
\divide\@tempdima\baselineskip
\@tempcnta=\@tempdima
\skip\@mpfootins = \skip\footins
\renewcommand{\@dotsep}{10000}

\newcommand\ps@myplain{
\pagenumbering{arabic}
\renewcommand\@oddfoot{\hfill-- \thepage\ --\hfill}
\renewcommand\@oddhead{}}
\let\ps@plain=\ps@myplain

\newcommand\ps@titlepage{\renewcommand\@oddfoot{}\renewcommand\@oddhead{}}

\numberwithin{equation}{section}

\renewcommand\section{\@startsection{section}{1}{\z@}%
                                   {-3.5ex \@plus -1.3ex \@minus -.7ex}%
                                   {2.3ex \@plus.4ex \@minus .4ex}%
                                   {\normalfont\large\bfseries}}
\renewcommand\subsection{\@startsection{subsection}{2}{\z@}%
                                   {-2.3ex\@plus -1ex \@minus -.5ex}%
                                   {1.2ex \@plus .3ex \@minus .3ex}%
                                   {\normalfont\normalsize\bfseries}}
\renewcommand\subsubsection{\@startsection{subsubsection}{3}{\z@}%
                                   {-2.3ex\@plus -1ex \@minus -.5ex}%
                                   {1ex \@plus .2ex \@minus .2ex}%
                                   {\normalfont\normalsize\bfseries}}
\renewcommand\paragraph{\@startsection{paragraph}{4}{\z@}%
                                   {1.75ex \@plus1ex \@minus.2ex}%
                                   {-1em}%
                                   {\normalfont\normalsize\bfseries}}
\renewcommand\subparagraph{\@startsection{subparagraph}{5}{\parindent}%
                                   {1.75ex \@plus1ex \@minus .2ex}%
                                   {-1em}%
                                   {\normalfont\normalsize\bfseries}}

\def\fnum@figure{\textbf{\figurename\nobreakspace\thefigure}}
\def\fnum@table{\textbf{\tablename\nobreakspace\thetable}}

\long\def\@makecaption#1#2{%
  \vskip\abovecaptionskip
  \sbox\@tempboxa{\small #1. #2}%
  \ifdim \wd\@tempboxa >\hsize
    \small #1. #2\par
  \else
    \global \@minipagefalse
    \hb@xt@\hsize{\hfil\box\@tempboxa\hfil}%
  \fi
  \vskip\belowcaptionskip}

\renewenvironment{thebibliography}[1]{%
\begin{oldthebibliography}{#1}%
\small%
\raggedright%
\setlength{\itemsep}{5pt plus 0.2ex minus 0.05ex}%
}%
{%
\end{oldthebibliography}%
}

\begin{document}


\title{\boldmath Why the entropy of spacetime is independent of species of particles ----- the
species problem}

\author[a]{Yu-Zhu Chen,}
\author[b,a]{Wen-Du Li,}
\author[a,c,1]{and Wu-Sheng Dai}\note{daiwusheng@tju.edu.cn.}


\affiliation[a]{Department of Physics, Tianjin University, Tianjin, 300350, P.R. China}
\affiliation[b]{Theoretical Physics Division, Chern Institute of Mathematics, Nankai University, Tianjin, 300071, P. R. China}
\affiliation[c]{LiuHui Center for Applied Mathematics, Nankai University \& Tianjin University, Tianjin 300072, P.R. China}








\abstract{The Hawking radiation emits all species of particles, but the
Bekenstein-Hawking entropy is independent of the number of the species of
particles. This is the so-called species problem --- a puzzling problem for a
long time. In this paper, we suggest a solution to this problem. A result of
the scheme is that the black hole atmosphere has a mass equaling $3/8$ mass of
a classical Schwarzschild black hole, which agrees with 't Hooft's brick wall model.
}


\maketitle
\flushbottom

By only taking a scalar field into account, one can obtain the whole
Bekenstein-Hawking entropy \cite{hooft1985quantum,solodukhin1995one}.
Nevertheless, the Hawking radiation emits all species of particles. What is
the role of other species of particles. It is hard to believe that only one
species of particle contributes to the entropy. This is the so called species
problem \cite{jacobson1994black,bekenstein1994we}.

In this paper, we suggest a solution to the species problem. A result of our
scheme is that the entropy of a spacetime is independent of the number of
species of particles. This means that even there are more than one species of
particles, even there exists some unknown species of particles, the entropy of
a spacetime is still the Bekenstein-Hawking entropy, as same as that
calculated from one single scalar field.

In our scheme, the black hole entropy is regarded as the entropy of the
atmosphere outside the horizon. The reason why the entropy of a spacetime can
be regarded as the entropy of its atmosphere is that the entropy of the
atmosphere is the entanglement entropy
\cite{wald2002thermodynamics,mukohyama1999aspects,susskind2005introduction},
and the entropy of a black hole can also be regarded as the entanglement
entropy
\cite{susskind1994black,dowker1994remarks,bombelli1986quantum,callan1994geometric}%
. More concretely, the entropy of the atmosphere was first calculated by t'
Hooft \cite{hooft1985quantum} in the brick wall model, and thereafter some
authors pointed out that the entropy of the atmosphere should be regarded as
the entanglement entropy
\cite{wald2002thermodynamics,mukohyama1999aspects,susskind2005introduction}.
In fact, the calculation of the entropy of a black hole is the same as the
entanglement entropy calculation \cite{wald2002thermodynamics}. The species
problem then arises in the entanglement entropy calculation.

The species problem has a direct bearing on the understanding of the spacetime
entropy. Much effort has been devoted to understanding the species problem.
The simplest scheme suggests that only a particular field contributes to the
spacetime entropy \cite{jacobson1994black,solodukhin1995one}. Based on the
Euclidean path integral method, Jacobson finds that the leading contribution
to the entropy in the effective action is the classical Hilbert-Einstein term
$\hbar a_{1}\int d^{4}x\sqrt{g}R$; by regarding the expansion coefficient
$\hbar a_{1}$ as the renormalized inverse gravitational constant $1/\left(
16\pi G\right)  $, the leading contribution is the same as the
Bekenstein-Hawking entropy \cite{jacobson1994black,larsen1996renormalization}.
Therefore, the entropy calculated by the Euclidean path integral is
independent of the number of species of particles. Frolov comes to a same
conclusion by a particular method \cite{frolov1995entropy}, who demonstrates
that the one-loop contribution to the entropy comparing with the tree-level
contribution is ignorable. A more extensively used scheme is to correct the
cutoff. When regarding the spacetime entropy as the entanglement entropy
\cite{wald2002thermodynamics,bekenstein1994we}, a short distance cutoff
$\Lambda$ must be brought in
\cite{bombelli1986quantum,frolov1993dynamical,hooft1985quantum}. By resetting
the cutoff to $\Lambda/\sqrt{N}$, where $N$ is the number of species of
particles, the entanglement entropy contributed by all species of particles
adds up to the Bekenstein-Hawking entropy \cite{dvali2008black}. Susskind
suggests a similar method, in which the gravitational constant $G$ is
renormalized to obtain the Bekenstein-Hawking entropy \cite{susskind1994black}%
. More discussions on the cutoff approach are also given by other authors
\cite{demers1995black,jacobson2007black,dvali2008black}.

Our scheme is as follows.

\textbf{Where there is a horizon there is an atmosphere.} The existence of the
Hawking radiation ensures that a spacetime with a horizon must have an
atmosphere consisting of the particles which are emitted out of the horizon
and fall back into the horizon in a dynamic equilibrium.

That is, a spacetime indeed consists of two indivisible parts: a bare
spacetime which is a classical spacetime solved from the Einstein equation and
its atmosphere which is a pure quantum effect. In other words, when talking
about a spacetime, we must simultaneously consider both the classical bare
spacetime and its atmosphere. These two parts are always concomitant with each
other from the birth of the spacetime. In our scheme, the bare spacetime and
its atmosphere are regarded as a single whole system.

A spacetime with an atmosphere\textbf{ }is indeed a dressed space consists of
a classical bare spacetime and a quantum atmosphere, just like a dressed
particle ----- a bare particle with a virtual-particle cloud. Nevertheless, it
should be emphasized that rather than the virtual-particle cloud of a
particle, the atmosphere of a spacetime consists of real particles.

\textbf{The entropy of a spacetime all comes from its atmosphere.} Since a
spacetime is indeed a classical bare spacetime with its atmosphere, the
entropy of a spacetime is then the sum of the entropy of the bare spacetime
and entropy of the atmosphere:
\begin{equation}
S=S_{bare}+S_{atmosphere}.
\end{equation}
The entropy $S=\ln\Omega$ with $\Omega$ the number of states. The bare
spacetime is regarded as a mechanical system and has only one certain state,
so it does not contribute to the entropy, i.e., $S_{bare}=\ln1=\allowbreak0$.
Consequently, the atmosphere contributes all the entropy of a spacetime:
$S=S_{atmosphere}$.

In a word, the entropy of a spacetime is just the entropy of the atmosphere of
the spacetime.

\textbf{All species of particles near the horizon behave like massless
particles.} The Hawking radiation emits all species of particles, massive and
massless. Nevertheless, all species of particles near the horizon behave like
massless particles.

We take a scalar particle near the horizon of a Schwarzschild spacetime as an
example to illustrate this.

The radial Klein-Gordon equation in a Schwarzschild spacetime with the
tortoise coordinate $r_{\ast}=r+r_{0}\ln\left(  r/r_{0}-1\right)  $ reads
\cite{susskind2005introduction}%
\begin{equation}
\left\{  -\frac{\partial^{2}}{\partial t^{2}}+\frac{\partial^{2}}{\partial
r_{\ast}^{2}}-\left(  1-\frac{r_{0}}{r}\right)  \left[  \frac{r_{0}}{r^{3}%
}-\frac{l\left(  l+1\right)  }{r^{2}}+m^{2}\right]  \right\}  R\left(
t,r\right)  =0, \label{B.3}%
\end{equation}
where $r_{0}=2M$ is the horizon radius with $M$ the mass of the classical bare
spacetime. Near the horizon, $r\rightarrow r_{0}$, the massive equation, eq.
(\ref{B.3}), reduces to a massless equation:
\begin{equation}
\left(  -\frac{\partial^{2}}{\partial t^{2}}+\frac{\partial^{2}}{\partial
r_{\ast}^{2}}\right)  R\left(  t,r\right)  =0. \label{B.4}%
\end{equation}
This shows that even a massive particle still has a behavior like a massless
particle near the horizon.

\textbf{The entropy of any species of massless particles has the same relation
to the internal energy.} Near the horizon,\ all particles behave like massless
particles. It can be shown that for both massless bosonic and massless
fermionic gases, i.e., ultrarelativistic bosonic and fermionic gases, the
relation between the entropy $S$ and the internal energy $U$ is
\cite{pathria2011statistical}%
\begin{equation}
S=\frac{4}{3}\beta U. \label{entropy}%
\end{equation}

It should be emphasized that near the horizon of the Schwarzschild spacetime,
the spacetime is almost flat \cite{susskind2005introduction}. This allows us
to apply the result of statistical mechanics in flat spacetime directly.

\textbf{The spacetime entropy is independent of the species of particles.}
Now, it is ready to describe our scheme.

It is already shown that no matter what species of particles are, bosonic or
fermionic, massive or massless, the entropies are always have the same
relation to the energies. Suppose that there are $N_{B}$ species of bosons and
$N_{F}$ species of fermions. The total energy of the atmosphere is, of
course,
\begin{equation}
U=\sum_{i=1}^{N_{B}}U_{i}^{Bose}+\sum_{i=1}^{N_{F}}U_{i}^{Fermi}.
\end{equation}
The entropy of the system, by eq. (\ref{entropy}), then reads%
\begin{align}
S  &  =\sum_{i=1}^{N_{B}}S_{i}^{Bose}+\sum_{i=1}^{N_{F}}S_{i}^{Fermi}%
\nonumber\\
&  =\sum_{i=1}^{N_{B}}\frac{4}{3}\beta U_{i}^{Bose}+\sum_{i=1}^{N_{F}}\frac
{4}{3}\beta U_{i}^{Fermi}\nonumber\\
&  =\frac{4}{3}\beta U.
\end{align}

This result tells us that \textit{no matter how many species of particles are
emitted, the entropy depends only on the total energy of particles rather than
the number of the species of particles}.

\textbf{The atmosphere accounts for }$\mathbf{3/11}$\textbf{ of the mass of a
Schwarzschild black hole. }For the Schwarzschild spacetime, we can arrive at
the Bekenstein-Hawking entropy%
\begin{equation}
S=4\pi M^{2}%
\end{equation}
so long as the total energy of the atmosphere is
\begin{equation}
U=\frac{3}{8}M.
\end{equation}
Note that for the Schwarzschild spacetime the reciprocal of the temperature
$\beta=8\pi M$.

Since the mass of the total system, $m_{total}$, in our scheme is the bare
mass, $m_{bare}=M$, plus the mass of the atmosphere, $m_{atmosphere}=\frac
{3}{8}M$, the ratio of the bare mass and the mass of the atmosphere is
$m_{bare}:m_{atmosphere}=8:3$, or, $m_{bare}:m_{total}=8:11$ and
$m_{atmosphere}:m_{total}=3:11$.

In a word, in this scheme the entropy is independent of the species of particles.

The key point of this scheme is to regard the atmosphere of a spacetime, which
is a pure quantum effect, as a part of the spacetime. In the scheme, a
spacetime consists of two parts: the classical bare spacetime and its
atmosphere. The entropy of the spacetime all comes from the atmosphere.
Considering that any particle near the horizon behaves like a massless
particle, we conclude that the atmosphere can be considered as an
ultra-relativistic gas. As a result, the entropy, regardless of the species of
particles, is proportional to the energy (mass) and is independent of the
species of particles.

Furthermore, since a spacetime is divided into a classical part (the solution
of the Einstein equation) and a quantum part (atmosphere), when studying the
quantum properties of a spacetime, one can only focus on the atmosphere of the
spacetime. That is to say, the property of a thermal field in a spacetime
background \cite{dai2010approach} reflects the quantum behavior of the spacetime.

A by-product of this scheme is that for a Schwarzschild black, the ratio of
the mass of the classical bare black hole and its atmosphere is $8:3$. Based
on the brick wall model, 't Hooft also showed that the energy of the bosonic
gas outside the Schwarzschild horizon is $\frac{3}{8}M$
\cite{hooft1985quantum}. Our result agrees with 't Hooft's result.

Finally, inspired by the hierarchy problem, we would like to suggest an
alternative scheme to understand the species problem. In a series of works
\cite{arkani1998hierarchy,dvali2010black,dvali2008black,dvali2013blacka,dvali2013blackb,dvali2009quantum}%
, the authors suggest a bound on the fundamental scale of the spacetime
imposed by the species of the particles in black hole physics. A modified
Planck length $l_{species}=\sqrt{N_{species}}l_{p}$ is introduced, where
$l_{p}$ is the Planck length and $N_{species}$ is the number of the species of
particles in the world \cite{dvali2013black}. The Bekenstein-Hawking entropy
of a black hole is $S_{BH}=\frac{1}{4}k_{B}A/l_{p}^{2}$. The entropy
contributed by one certain species of particles is $S_{species}=\frac{1}%
{4}k_{B}A/l_{species}^{2}$, in which the Planck constant $l_{p}$ is replaced
by the modified Planck length $l_{species}$. If there are $N$ species of
particles in the world, the entropy of a black hole is the sum of the
contributions of all species of particles. We then have $S_{BH}=NS_{species}%
=\frac{1}{4}k_{B}A/l_{p}^{2}$. This result tells us that despite how many
species of particles in the world, the entropy of a black hole is still the
Bekenstein-Hawking entropy. In other words, the entropy is independent of the
species of particles.


\acknowledgments

We are very indebted to Dr G. Zeitrauman for his encouragement. This work is supported in part by NSF of China under Grant Nos. 11575125 and 11675119.










\providecommand{\href}[2]{#2}\begingroup\raggedright\endgroup


\end{document}